\documentclass[prl,twocolumn,amsmath,amssymb,aps,a4paper,floatfix,10pt,superscriptaddress]{revtex4-1} 
\pdfoutput=1
\usepackage{graphicx}
\usepackage{bm}
\usepackage{amssymb,amsmath}
\usepackage{natbib}
\usepackage{verbatim}
\usepackage[utf8]{inputenc}
\usepackage[T1]{fontenc}

\newcommand{\fig}[1]{Fig.~\ref{#1}}

\begin{document}

\title{Laser cooling and control of excitations in superfluid helium}

\author{G. I. Harris} \thanks{These authors contributed equally to this work.} \affiliation{Centre for Engineered Quantum Systems, School of Mathematics and Physics, University of Queensland, Brisbane, QLD 4072, Australia.} 
\author{D. L. McAuslan} \thanks{These authors contributed equally to this work.} \affiliation{Centre for Engineered Quantum Systems, School of Mathematics and Physics, University of Queensland, Brisbane, QLD 4072, Australia.} 
\author{E. Sheridan} \affiliation{Centre for Engineered Quantum Systems, School of Mathematics and Physics, University of Queensland, Brisbane, QLD 4072, Australia.}
\author{Y. Sachkou} \affiliation{Centre for Engineered Quantum Systems, School of Mathematics and Physics, University of Queensland, Brisbane, QLD 4072, Australia.}
\author{C. Baker} \affiliation{Centre for Engineered Quantum Systems, School of Mathematics and Physics, University of Queensland, Brisbane, QLD 4072, Australia.}
\author{W. P. Bowen} \affiliation{Centre for Engineered Quantum Systems, School of Mathematics and Physics, University of Queensland, Brisbane, QLD 4072, Australia.}


\date{\today}

\maketitle


\textbf{
Superfluidity is an emergent quantum phenomenon which arises due to strong interactions between elementary excitations in liquid helium. These excitations have been probed with great success using techniques such as neutron and light scattering\cite{Tilley_Book, Bramwell_natmat14}. However measurements to-date have been limited, quite generally, to average properties of bulk superfluid\cite{Tilley_Book} or the driven response far out of thermal equilibrium\cite{Gorter_Book, Ellis_PRB89, Hoffmann_JLTP04, DeLorenzo_NJP14}. Here, we use cavity optomechanics\cite{Chan11_Nat, Brooks12_Nat, Palomaki_Sci13, Schreppler_Sci14} to probe the thermodynamics of superfluid excitations in real-time. Furthermore, strong light-matter interactions allow both laser cooling and amplification of the thermal motion. This provides a new tool to understand and control the microscopic behaviour of superfluids, including phonon-phonon interactions\cite{Tilley_Book}, quantised vortices\cite{Penanen_JLTP02} and two-dimensional quantum phenomena such as the Berezinskii-Kosterlitz-Thouless transition\cite{Kosterlitz73_JPhysCSS}. The third sound modes studied here also offer a pathway towards quantum optomechanics with thin superfluid films, including femtogram effective masses, high mechanical quality factors, strong phonon-phonon and phonon-vortex interactions, and self-assembly into complex geometries with sub-nanometre feature size.
}

Elementary excitations, in the form of phonons and rotons, are fundamental to both the macroscopic and microscopic quantum behaviour of superfluid helium-4, including phenomena such as dissipation\cite{Hoffmann_JLTP04,Ellis_PRB89}, quantum turbulence\cite{Barenghi14_PNAS} and quantum phase transitions\cite{Bishop_PRL78}
Techniques to probe such excitations have been crucial to our understanding of superfluids since the 1960s\cite{Tilley_Book}. For instance, neutron and light scattering\cite{Tilley_Book, Bramwell_natmat14, Pike_JPhysC70} allow the dynamic structure factor to be determined, which quantifies the dispersion relation, as well as the mean occupancy and correlations. However, such techniques are slow compared to the characteristic dissipation rate of the excitations, constraining them to average thermodynamical properties of the superfluid and prohibiting real-time measurement and control. Real-time measurements have only previously been performed by applying an external driving force to individual modes in superfluid acoustic resonators\cite{Brooks_PRL78, Hoffmann_JLTP04, DeLorenzo_NJP14}. This excites them far out of thermal equilibrium and constrains measurements to coherent dynamics. 

In cavity optomechanics, the coupling between optical and mechanical degrees-of-freedom, and therefore measurement rate, is greatly enhanced by the presence of a high quality optical cavity. This has enabled the demonstration of a range of quantum behaviour \cite{Brooks12_Nat, Verhagen12_Nat, Palomaki_Sci13}; measurement precision approaching the fundamental limit set by quantum uncertainty\cite{Schreppler_Sci14}; and precision sensors of mass, acceleration and magnetic fields\cite{Metcalfe_APR14}. Excitations in superfluids have recently been identified as an attractive mechanical degree-of-freedom\cite{DeLorenzo_NJP14, Agarwal_PRA14}; introducing unique features such as viscosity that approaches zero at absolute zero, quantized rotational motion and vortices\cite{Penanen_JLTP02}, and strong phonon-phonon interactions\cite{Tilley_Book}.
In the only previous experiment a pressure wave in bulk helium acts as a gram-scale resonator\cite{DeLorenzo_NJP14}, with the combination of high mechanical quality factor and mass providing a path towards ultra-precise inertial sensors. However, the comparatively large mass presents significant challenges for the observation or control of thermal excitations, and the manifestation of quantum effects.

\begin{figure*}
\begin{center}
\includegraphics[width=0.95\textwidth]{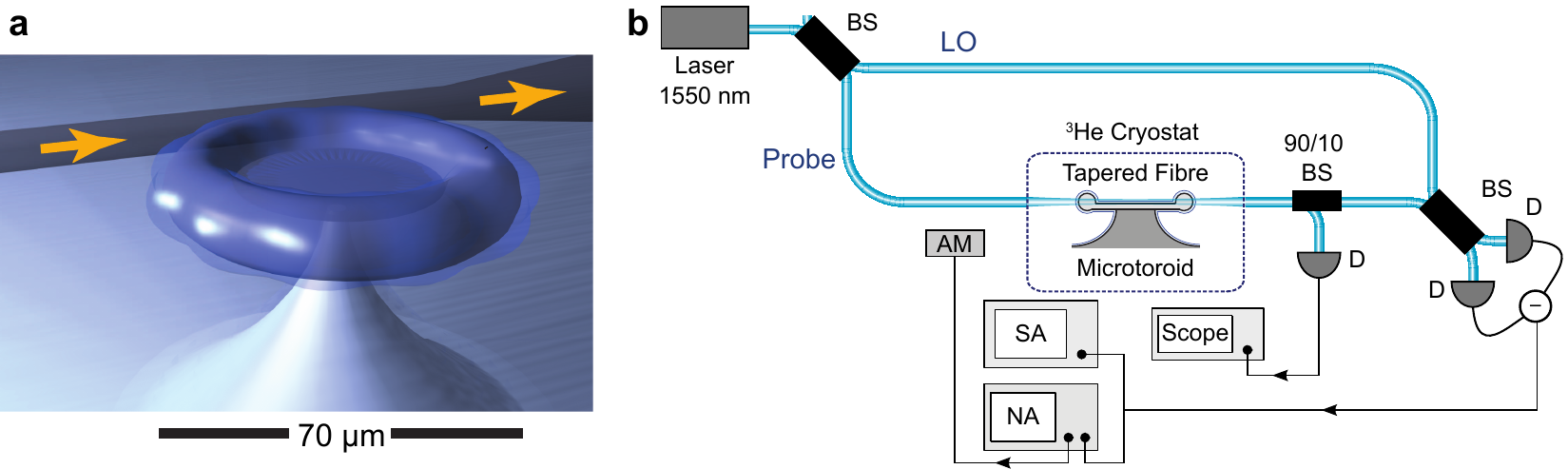}
\end{center}
\caption{\label{fig:expsetup} \textbf{Optomechanics with superfluid helium films.} (\textbf{a}) Illustration of third sound waves of a superfluid helium film coating a microtoroid.  The third sound oscillations have phase velocity $C_3 = \sqrt{3\alpha_{vdw} d^{-3}}$, where $\alpha_{vdw} = 2.65 \times 10^{21}\,\rm nm^5s^{-2}$ is the van der Waals coefficient for silica. 
(\textbf{b}) Shot-noise limited homodyne detection of superfluid helium optomechanics. A tapered fibre-coupled microtoroid (major diameter = 70~$\mu$m, minor diameter = 7~$\mu$m) is mounted inside the sample chamber of a helium-3 cryostat. Helium-4 gas is injected into the sample chamber at a pressure of 69~mTorr (at 2.8~K). AM - amplitude modulator, BS - beamsplitter, D - photodetector, NA/SA - network/spectrum analyser. The 90/10 BS probes the optical transmission through the microtoroid. At the base temperature of the cryostat the superfluid film is roughly 10~nm thick. 
}
\end{figure*}

Here we propose and utilise an alternative approach to superfluid optomechanics based on femtogram films of superfluid helium condensed on the surface of a microscale whispering-gallery-mode resonator (Fig.~\ref{fig:expsetup}a). Superfluid films form naturally on surfaces due to the combination of ultralow viscosity and attractive van der Waals forces. Excitations in such films, known as third sound\cite{Atkins_PR59, Ellis_PRB89, Shirron_PRL91, Hoffmann_JLTP04}, manifest as perturbations to the thickness with the restoring force provided by the van der Waals interaction. The physical structure of the resonator provides a template for the self-assembling film, acting to confine third sound modes at the microscale in two dimensions, with film thickness defining the third dimension. 
Optomechanical coupling is realized via the optical evanescent field\cite{Anetsberger_NatPhys09}, with the film being naturally located in the region of maximum field strength.
Compared with previous third sound experiments that use centimetre-scale enclosures and capacitive measurements, this architecture enables three orders-of-magnitude reduction in mechanical mode volume combined with greatly enhanced readout sensitivity\cite{Ellis_PRB89, Shirron_PRL91, Hoffmann_JLTP04}.

\begin{figure*}
\begin{center}
\includegraphics[width=1\textwidth]{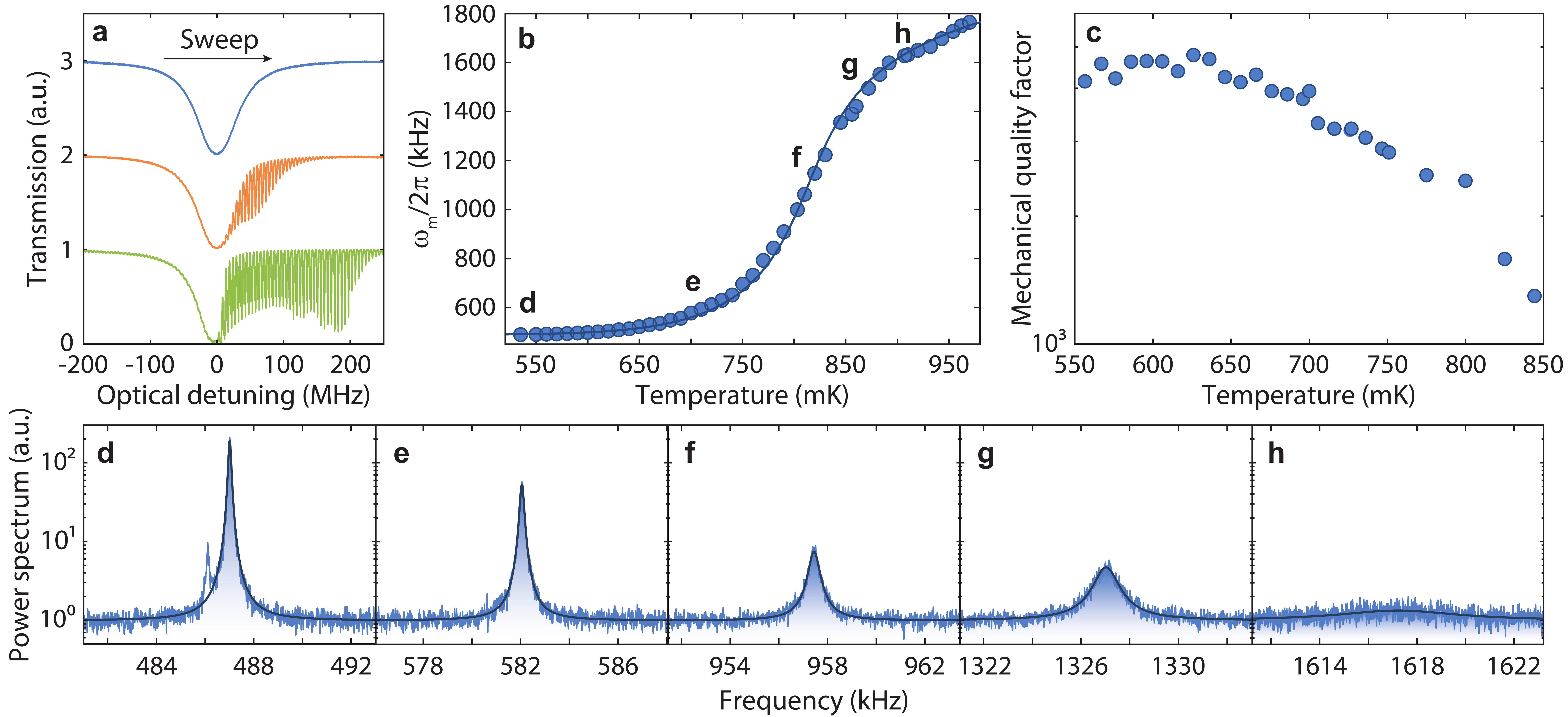}
\end{center}
\caption{\label{fig:figure2} \textbf{Superfluid helium mechanics.}
(\textbf{a}) Observation of superfluid oscillations as the cryostat cools to base temperature, while the laser frequency is scanned and optical resonance tracked. Blue, orange, and green traces are offset vertically and were respectively taken at 3~K, 1~K, and 0.6~K.
(\textbf{b}) Resonance frequency of a particular superfluid mode versus cryostat temperature, with a final frequency of $482$~kHz. The solid line is a theoretical fit obtained by modelling the condensation of the helium gas.
(\textbf{c}) Mechanical quality factor of the superfluid mode versus cryostat temperature. Note above $850$~mK the signal to noise ratio was too low to accurately measure the quality factor.
(\textbf{d-h}) Mechanical spectra of the superfluid mode at 530, 700, 800, 850, and 900~mK (from left to right). These measurements were performed with the laser coupled to a microtoroid optical mode at $\lambda = 1555.1$~nm with optical decay rate $\kappa/2\pi = 22.3$ MHz.
}
\end{figure*}

To experimentally realize thin-film superfluid optomechanics, a fibre coupled microtoroidal resonator is placed in a low pressure helium-4 gas environment within a helium-3 cryostat (Fig.~\ref{fig:expsetup}b). At the pressures used the helium gas transitions directly to the superfluid state at approximately 1 kelvin, avoiding the normal fluid phase. Van der Waals forces then coat the surface of the sample chamber with a film of superfluid helium. A typical optical resonance lineshape is seen at temperatures above the gas-to-superfluid transition (\fig{fig:figure2}a (blue)). Below the transition, unstable oscillations appear on the blue detuned side of resonance (\fig{fig:figure2}a (orange, green)), characteristic of optomechanical parametric instability. This is observed with as little as 40~nW of optical power, over one hundred times lower than would be expected from a microtoroid mechanical mode\cite{Harris12_PRA}. 

To characterize the mechanical response of the superfluid film we lock the laser to an optical resonance of the microtoroid. Amplitude modulation allows coherent driving of the motion of the superfluid via photothermal actuation\cite{ Restrepo_CRP11}. This arises due to the superfluid fountain effect where superfluids flow towards a localised heat source. The modulation is monitored via the phase quadrature of the optical field. Sweeping the modulation frequency a dense spectrum of mechanical modes is observed (see Supplementary Information), ranging in frequencies from 10~kHz to 5~MHz consistent with third sound waves confined to length scales on the order of the microtoroid dimensions. The modes tend to exist in near degenerate pairs, separated in frequency by several hundred hertz. We attribute this to breaking of cylindrical symmetry due to scattering by a defect on the microtoroid surface. As the cryostat cools the frequency of the superfluid modes is observed to decrease (Fig.~\ref{fig:figure2}b). This occurs due to increased condensation of helium into the superfluid film, with thickening of the film resulting in a weaker van der Waals mediated restoring force. The low temperature plateau in frequency, evident in Fig.~\ref{fig:figure2}b, occurs when the majority of gaseous helium in the sample chamber has been condensed into the superfluid phase, resulting in a mechanical frequency that can be precisely tuned by injecting or evacuating helium gas.

When coupled to a thermal bath at temperature $T$, the root-mean-square motional amplitude of an oscillator is given by equipartition to be $\delta x = \sqrt{k_B T/k}$ where $k$ is the spring constant of the oscillator. The motion decorrelates over a characteristic timescale of $2\pi/\Gamma_m$, where $\Gamma_m$ is the oscillator decay rate. Only if measurement precision better than $\delta x$ is achieved within this time scale is it possible to track the thermally driven trajectory of the oscillator in phase space. This allows thermodynamical fluctuations to be studied and controlled in real-time, and the fundamental thermomechanical noise floor of force and inertial sensing to be reached.
While many techniques have been developed to probe the thermal properties of superfluid helium, it has proved difficult to achieve this regime.
For instance, the light scattering measurements in Ref.~\cite{Pike_JPhysC70} averaged photocounts for around 30 minutes to retrieve the thermal motion spectrum of first sound waves, while the recent work of De Lorenzo and Schwab\cite{DeLorenzo_NJP14} remains six orders of magnitude away from resolving the thermal motion.

\begin{figure}
\begin{center}
\includegraphics[width=0.40\textwidth]{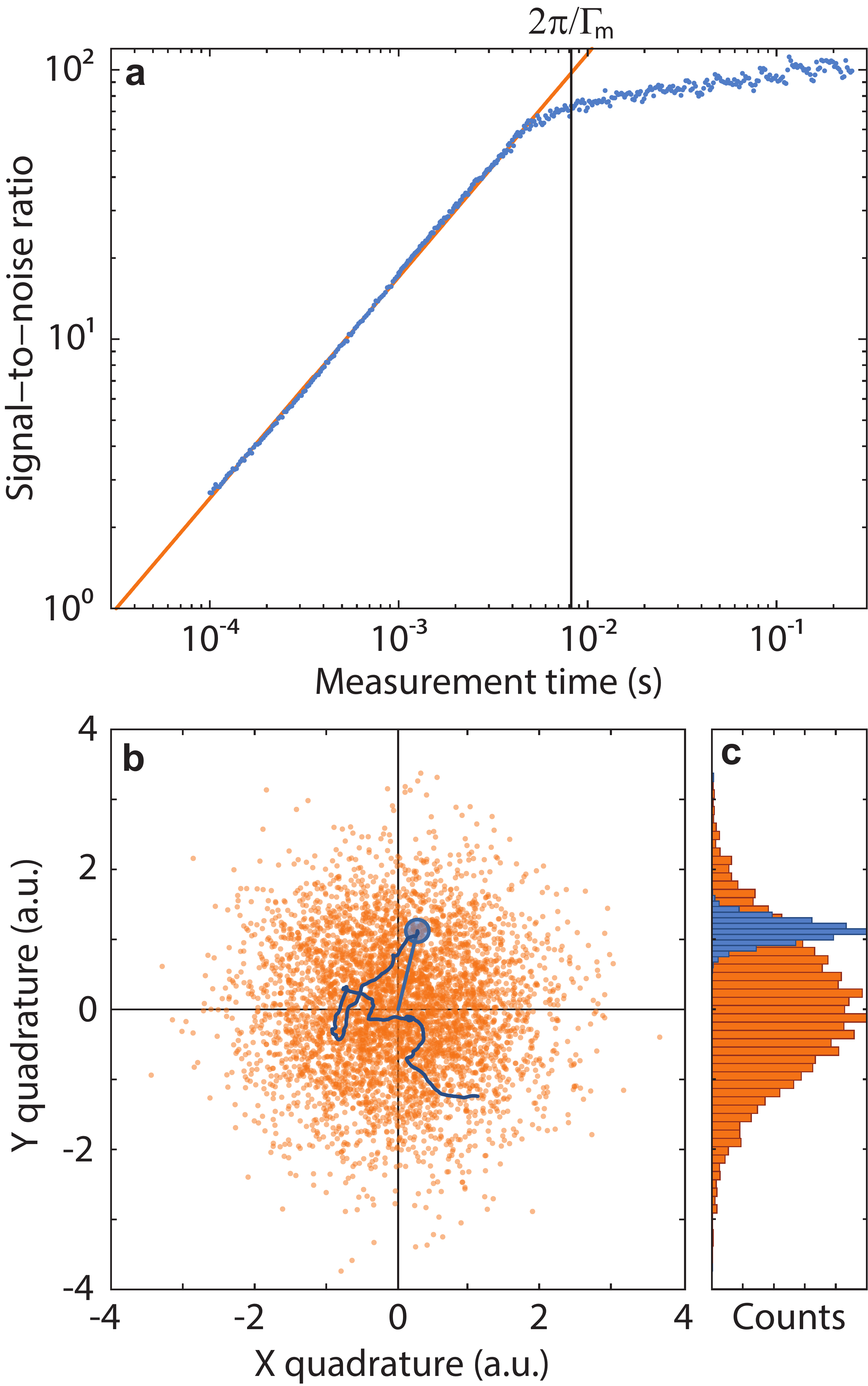}
\end{center}
\caption{\label{fig:realtime} \textbf{Realtime measurements of superfluid motion.}
(\textbf{a}) Signal-to-noise ratio of mechanical spectra of the 482~kHz mode as the measurement time is decreased. Measurements were made at the base temperature of our cryostat (530~mK).
(\textbf{b}) Thermal motion of the superfluid mode in phase space. Each orange point is the position of the oscillator, measured in a time $2\pi/4\Gamma_m$. 
The X and Y quadratures of the oscillator position are acquired by optimally Wiener filtering the homodyne signal, then mixing this down to DC.
$X = W\left(i\left(t\right))\right)\cos\left(\omega_m t\right)$, $Y = W(i(t))\sin(\omega_m t)$, where $i(t)$ is the homodyne signal and $W$ is the Wiener filter.
The blue circle represents the uncertainty of an individual measurement, defined as the standard deviation of the shot-noise in the measurement time. The dark blue line shows an example trajectory of the oscillator obtained by making successive measurements, tracking its motion in real-time over a period of 15~ms.
(\textbf{c}) Histogram of the position of the superfluid (orange) and the measurement noise (blue) shown in (\textbf{b}). Both the superfluid position and measurement uncertainty are normally distributed. Statistics were built up by binning the result of 4700 individual measurements taken over 5~s of data acquisition.
%
}
\end{figure}

The reduced mode volume and strong evanescent optomechanical coupling achieved in our architecture combine to greatly enhance the capacity to resolve thermodynamical fluctuations.
To test whether real-time measurements are possible we perform homodyne-based phase measurement.
Spectral analysis on a high quality third sound mode at 482~kHz reveals a thermomechanical noise peak characteristic of mechanical oscillations (see Figs.~\ref{fig:figure2}d-h).
%
%
It was observed that the mechanical quality factor increases substantially with decreasing temperature (\fig{fig:figure2}c) consistent with previous observations attributed to phonon-vortex interactions\cite{Penanen_JLTP02, Hoffmann_JLTP04}, with the dissipation rate reaching a minimum of $\Gamma_m/2\pi = 106\,\rm Hz$ at 530~mK. 
To determine the minimum measurement time required to resolve the thermomechanical motion, the signal-to-noise ratio (SNR) of this peak was determined as a function of measurement duration (Fig.~\ref{fig:realtime}a).
At measurement times greater than $2\pi/\Gamma_m$ the SNR is relatively constant, as expected, and reaches a value as high as $20.5$~dB.
When the measurement time is reduced below $2\pi/\Gamma_m$ the SNR drops linearly. Extrapolating to a SNR of one we find that the superfluid motion is resolvable for measurement times as low as 32~$\mu$s; a factor of 260 times shorter than $2\pi/\Gamma_m$ (8.2~ms), and therefore sufficient to track the superfluid motion in real time (Fig.~\ref{fig:realtime}a).
To demonstrate this capability, we monitored the evolution of the superfluid mode in phase space as a function of time from a sequence of measurements each having 2~ms duration (Fig.~\ref{fig:realtime}b). As can be seen, it was possible to track the thermal trajectory of the oscillator with precision a factor of 6.2 below the thermomechanical noise (Fig.~\ref{fig:realtime}c).

\begin{figure*}[t]
\begin{center}
\includegraphics[width=0.75\textwidth]{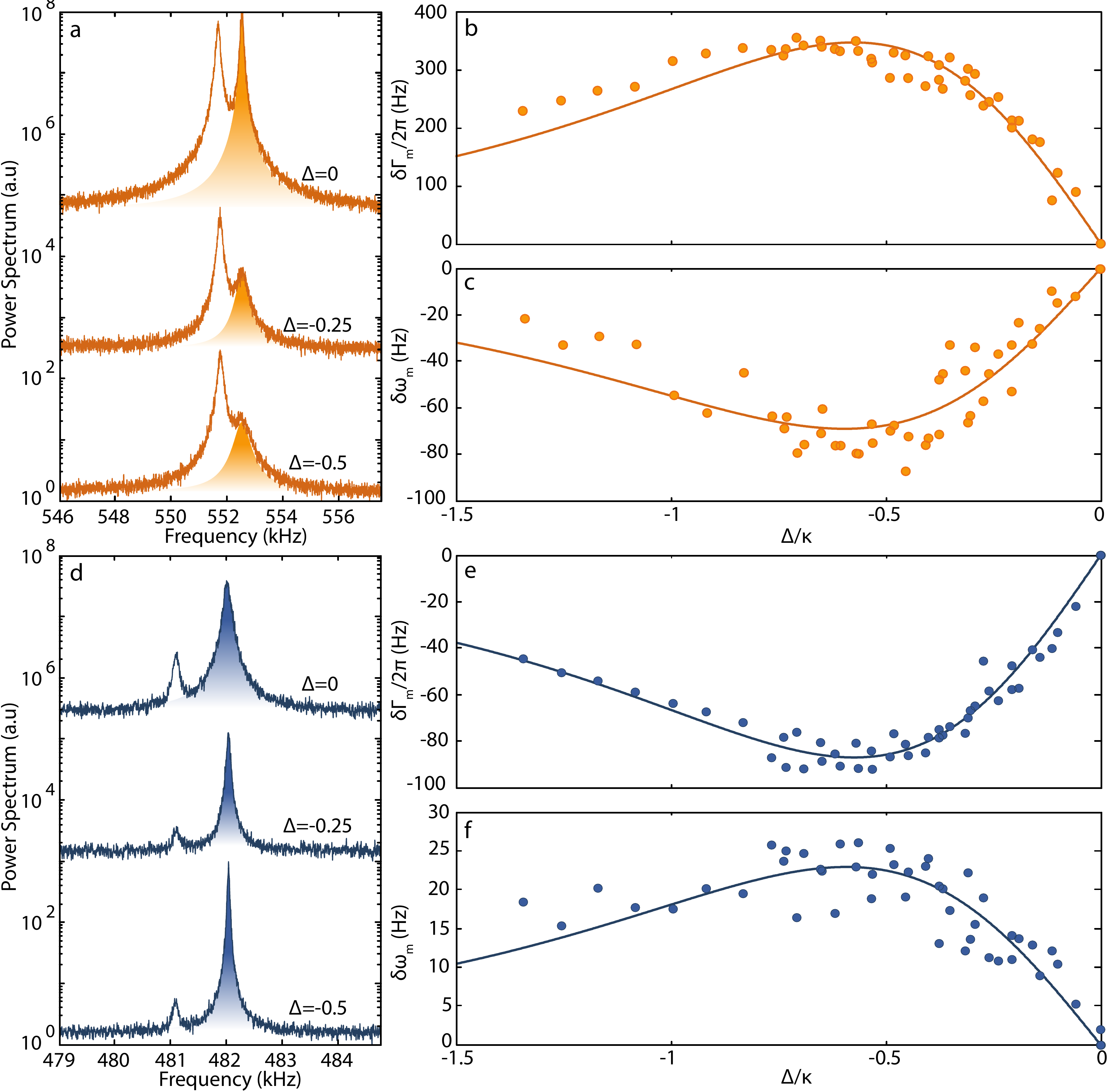}
\end{center}
\caption{\label{fig:heatandcool} \textbf{Optomechanical heating and cooling of two third sound modes.} Detuning the optical field results in cooling (\textbf{a}) or heating (\textbf{d}) of the superfluid excitations. The laser is red-detuned with respect to the cavity for both the $552.5$~kHz mode and the $482$~kHz mode. Whether the mode experiences heating or cooling depends on the relative magnitude of the radiation pressure force and the sign of the photothermal constant ($\beta$), which itself depends on the optical and mechanical mode. 
(\textbf{a}) Spectra of two closely spaced mechanical modes around $552.5$~kHz with varying optical detuning. The shaded mode experiences photothermal induced broadening ($\beta>0$), with a mechanical linewidth of $\Gamma_m/2\pi = 77$~Hz ($\Delta/\kappa = 0$), $\Gamma_m/2\pi = 334$~Hz ($\Delta/\kappa = -0.25$) and $\Gamma_m/2\pi = 457$~Hz ($\Delta/\kappa = -0.5$). 
The difference in photothermal response between adjacent modes can be explained by local heating from optical absorption by the defect responsible for the observed mode splitting, since the defect can be expected to reside at an antinode of the higher frequency mode, and a node of the lower frequency mode.
(\textbf{b}-\textbf{c}) Change in mechanical damping rate and mechanical resonance frequency as a function of detuning for the $552$~kHz mode. A maximum relative frequency shift of $\delta\omega_m/2\pi=-60$~Hz is measured at $\Delta=-0.60 \kappa$. Solid line: theoretical fit to photothermal broadening. 
(\textbf{d}) Spectra of two closely spaced mechanical modes around $482$~kHz with varying optical detuning. The shaded mode experiences photothermal induced narrowing ($\beta<0$), with a mechanical linewidth of $\Gamma_m/2\pi = 115$~Hz ($\Delta/\kappa = 0$), $\Gamma_m/2\pi = 63$~Hz ($\Delta/\kappa = -0.25$) and $\Gamma_m/2\pi = 36$~Hz ($\Delta/\kappa = -0.5$) . 
(\textbf{e}-\textbf{f}) Change in mechanical damping rate and mechanical resonance frequency as a function of detuning for the $482$~kHz mode. A maximum relative frequency shift of $\delta\omega_m/2\pi=23$~Hz is measured at $\Delta =-0.60\kappa$. Solid line: theoretical fit to photothermal narrowing. Traces in \textbf{a} and \textbf{d} are offset for clarity. All detuning measurements were taken with $200$~nW of launched optical power.
}
\end{figure*}

The ability to resolve the thermal motion of the superfluid allows thermometry to be performed on superfluid modes. 
Locking the laser to the cavity resonance, it is possible to probe the effect of the optical field on the superfluid modes in the absence of dynamical optomechanical backaction. Spectral analysis of the modes in this regime reveals an increase in mode temperature and linewidth with laser power. 
Thermometry of a microtoroid mechanical mode shows that this is not due to bulk optical heating (see Supplementary Information). This is further evidenced by the scaling of both temperature and linewidth with optical power, which was observed to be sub-linear. Nonlinear scaling of this form has been observed in other optomechanical systems\cite{Meenehan_PRA14}, and can be explained by the presence of an independent bath, coupled to the superfluid mode, and driven out of equilibrium by the optical field (see Supplementary Information for theoretical models and experimental data).

To experimentally probe the effect of dynamical backaction we red-detune the optical field from the cavity resonance, observing modifications to the mechanical resonance frequency and dissipation rate (Fig. \ref{fig:heatandcool}). The combination of radiation pressure and photothermal forces provides a mechanism to both cool and heat the superfluid excitations (Fig. \ref{fig:heatandcool}a,d). 
It was found that the dynamical backaction is dominated by photothermal forces\cite{Restrepo_CRP11}. The strength of these forces is determined by the spatial overlap between optical and mechanical modes. Consequently, mechanical modes of similar frequency may experience vastly different photothermal forces. This is evident in Fig.~\ref{fig:heatandcool}a where the shaded mode at $552.5$~kHz has a significantly stronger response compared with the adjacent mechanical mode. 
%

The photothermally induced changes in damping rate and resonance frequency of the $552.5$~kHz mode are shown in Fig. \ref{fig:heatandcool}b,c. The linewidth broadens from $\Gamma_0/2\pi = 115$~Hz on-resonance to $\Gamma_\Delta/2\pi = 464$~Hz at a detuning of $\Delta=-0.58 \kappa$.
This behaviour agrees well with photothermal theory with a positive photothermal coefficient (see Supplementary Information), as seen by the fit in Fig.~\ref{fig:heatandcool}a-c. 
From the change in damping rate we find that detuning cools the third sound mode by a maximum factor of $\frac{T_{\Delta}}{T_0} = \frac{\Gamma_{0}}{\Gamma_{\Delta}} = 0.25$.
%

In contrast to the $552.5$~kHz mode, dynamical backaction experiments on the $482$~kHz mode shows linewidth narrowing and spring stiffening with red-detuning (Fig.~\ref{fig:heatandcool}d), indicating that here the photothermal coefficient is negative, opposing the direction of radiation pressure. At a detuning of $\Delta=-0.58 \kappa$ the linewidth has narrowed to $\Gamma_\Delta/2\pi =49$~Hz from $\Gamma_0/2\pi =137$~Hz on-resonance.  This corresponds to an increase in oscillator temperature of $\frac{T_{\Delta}}{T_0} = 2.8$. 
Again good agreement with theory is achieved as seen by the fit in Fig. \ref{fig:heatandcool}e,f (see Supplementary Information).
%

Compared to other cavity optomechanics experiments that utilize photothermal coupling, the photothermal effect observed here is exceptionally fast due to the high thermal conductivity of superfluid helium\cite{Tilley_Book}. The characteristic time constant $\tau_t$ can be deduced from the functional form of the photothermal response seen in Fig. \ref{fig:heatandcool}b,c and Fig. \ref{fig:heatandcool}e,f and is found to be approximately $\tau \approx600$~ns.


Quasiparticles are believed to play a central role in both the microscopic and macroscopic behaviour of superfluid helium. However, our understanding remains incomplete, particularly in two-dimensional films. For instance, there remains significant debate about how third sound dissipates, with phonon-vortex interactions thought to play a crucial role\cite{Hoffmann_JLTP04}. Phonon-vortex interactions are also important for as-yet untested predictions such as the formation of Onsager vortices as two-dimensional superfluids evolve\cite{Simula_PRL14} and the decay of two-dimensional superfluid turbulence at zero temperature\cite{Kozik_PRB05,Davis_PB00}. 
%
By providing the capacity in thin superfluid films to both resolve and control thermodynamical motion, and with orders-of-magnitude smaller superfluid volumes than has previously been possible, this paper provides a new tool to study problems of this kind.

In general, interactions are enhanced as systems become increasingly confined. In superfluid helium, for instance, the single phonon-vortex interaction energy scales as the inverse-square of the confining length-scale. In our devices, this energy is approximately $10$~Hz, four orders-of-magnitude larger than any previous experiment\cite{Ellis_PRB89}, and only a factor of ten smaller than the phonon dissipation rate. As a consequence, our devices offer the prospect to resolve vortex-phonon dynamics in real-time with as few as ten unpaired vortices. Combined with atomically smooth surface finish that reduces the occurrence of pinned-vortices, this should allow a new approach to the study of third sound dissipation mechanisms.

Third sound dissipation also provides a method to probe quantum phase transitions such as the Berezinskii-Kosterlitz-Thouless (BKT) transition\cite{Kosterlitz73_JPhysCSS}. The BKT transition, while well studied in weakly interacting condensates, is not fully understood in dense, strongly interacting systems such a superfluid helium-4. Phonon-vortex interactions play an important role near the phase transition\cite{Kosterlitz73_JPhysCSS}. 
The enhanced confinement and measurement precision reported here will allow BKT to be probed with precision beyond that available with current technologies
Furthermore, with a factor of three reduction in diameter, it should be possible to reach the strong coupling regime, where the single phonon-vortex interaction energy is larger than both phonon and vortex dissipation rates, with phonons and vortices hybridizing into a new form of quasiparticle. This offers the prospect of studying a new, previously unexplored regime of BKT physics. Such physics would be further augmented by the capacity to laser control and cool individual phonon-modes demonstrated here. While laser cooling is routinely achieved in solid- and gas-phase systems such as cold atom physics, ion trapping, atomic clocks and optomechanics, it has not previously been demonstrated in superfluid helium, or indeed in any liquid.

We finally observe that superfluid films may open new regimes of cavity optomechanics. Hybridized phonon-vortex modes may allow experiments where the optical field is strongly coupled to an inherently quantized degree of freedom. Furthermore, third sound modes have been observed in helium films thinner than a single atomic layer\cite{Shirron_PRL91}. This regime is inherently highly nonlinear, with mechanical zero point fluctuations large compared to the film thickness. In sensing applications, atom interferometry with superfluid helium enables precise force and inertial sensing. The ability to resolve the thermomechanical motion of the fluid demonstrated here could enable superfluid force and inertial sensors that operate at the thermal noise limit, a capability which until now has not been possible.

\end{document}


\title{Supplementary Information}
\maketitle

\section*{Two fluid model and third sound}
Mechanical motion of liquid helium is described by the two-fluid model, where the liquid is composed of a normal fluid and a superfluid component that intermingle without any viscous interaction \cite{Donnelly09_PT, Tilley_Book}. This model gives several types of motion such as first and second sound. The mechanical motion studied here is that of third sound in superfluid helium-4, which refers to surface waves occurring on nanoscale films ($\sim10$~nm thick) \cite{Atkins_PR59, Everitt_PRL62, Everitt_PR64, Donnelly_JPC98}. In this case the film is so thin that the normal fluid component is clamped to the substrate, leaving only the superfluid component free to oscillate perpendicular to the surface. Third sound is completely analogous to classical waves on water; however, the presence of superfluidity is necessary because the waves would be quickly attenuated if it were a normal fluid. 

For third sound waves the restoring force is provided by the van der Waals interaction of the helium with the substrate. The resulting mechanical oscillations have a velocity ($c_s$) that is highly dependent on the thickness of the superfluid layer \cite{Atkins_PR59}:

\begin{equation}\label{eq:thirdsound} 
c_s = \sqrt{3 \frac{\rho_s}{\rho}\frac{\alpha_\mathrm{VDW}}{d^3}},
\end{equation}

where $\rho_s/\rho$ is the ratio of superfluid to total fluid density, and $\alpha_\mathrm{VDW}$ is the van der Waals coefficient ($\alpha_\mathrm{VDW} = 2.65 \times 10 ^{21}$~nm$^5$s$^{-2}$ for helium on a  silica substrate~\cite{Scholtz_PRL74}).

\section*{Theoretical Treatment}
Here we will theoretically explore the situation of generalized optomechanical coupling with both radiation pressure and photothermal forces\cite{Metzger_Nat04, Restrepo_CRP11, Usami_Nat12}. As highlighted in Ref. \cite{Restrepo_CRP11} the photothermal effect is not well described by an energy conserving Hamiltonian formalism owing to its inherently dissipative nature. This is in stark contrast to the well known dispersive optomechanical Hamiltonian, where energy and momentum are conserved by a reversible interaction between optical and mechanical degrees of freedom.

Using an approach similar to that taken in Ref. \cite{Restrepo_CRP11} the dispersive optomechanical Hamiltonian is represented in the Heisenberg picture including coupling to a ``bath'' of resonators that introduces both fluctuations and dissipation into the system. Similarly the photothermal effect is introduced as a fluctuating force with temporal correlations generated by the thermal response to the random absorption of photons. The quantum Langevin equation for the oscillator position $x(t)$ is:
\begin{eqnarray}
m_\text{eff}\left[\ddot{x}(t)+\Gamma_m\dot{x}(t)+\omega^2_m x(t)\right] &=& F_{RP}+F_{PT}+F_{th}\\ \label{eq:SF_x}
&=& \hbar g\left[ a(t)^{\dagger}a(t) + \frac{\beta A}{\tau_{t}}  \int_{-\infty}^{t} \mathrm{d}u \: e^{- \frac{t-u}{\tau_t}}a(u)^\dagger a(u) \right] \\  \nonumber
&&+ \sqrt{2\Gamma_m m_\text{eff} k_b T} \: \xi(t) 
\end{eqnarray}
where $A$ is the absorption coefficient given by the ratio of absorbed to circulating optical power and $\tau_t$ is the thermal response time. The dimensionless parameter $\beta$ quantifies the relative strength of the photothermal process over the radiation pressure such that $F_{PT} = \beta A F_{RP}$. As will be discussed later the absolute value and even the sign of $\beta$ is strongly dependent on the spatial overlap of the optical and mechanical mode. The intracavity field is dispersively coupled to the mechanical motion so the quantum Langevin equation for the annihilation operator $a(t)$ is :

\begin{eqnarray}
\dot{a}(t) &=& -\left[ \kappa - i\left(\Delta^0+gx(t) \right)\right]a(t) + \sqrt{2\kappa_{in}}a_{in}(t)
\label{eq:SF_a}
\end{eqnarray}
where the optomechanical coupling is $g$, $\Delta^0$ is the optical detuning, and the cavity decay is $\kappa =\kappa_{in} + \kappa_0$. It is useful to note that photothermal backaction is enabled by the temporal delay of the thermal response $\tau_t$ in (\ref{eq:SF_x}), and not the cavity induced delay as in standard dispersive optomechanics. As such, for photothermal cooling the condition of being ``resolved'' considers only the thermal response time, hence relaxing the stringent requirements on minimizing the optical decay rate. 

To expand the equations of motion into linear and nonlinear components the position and intracavity field annihilation operator are expressed as a coherent amplitude with quantum fluctuations i.e. $x=\bar{x}+\delta x(t)$, $a(t) = \alpha + \delta a(t)$ and $a_{in}(t) = \alpha_{in} + \delta a_{in}(t)$. Here we will only consider a linear photothermal and radiation pressure interaction so the nonlinear terms in (\ref{eq:SF_x}) and (\ref{eq:SF_a}) become: 

\begin{align}
%
iga(t)x(t) &= ig\left(\alpha\bar{x} + \delta x(t)\alpha +\delta a(t)\bar{x}\right) \\
\hbar g a^\dagger(t) a(t) &= \hbar g \left(|\alpha|^2 + 2\alpha\delta X^+(t)\right) 
\end{align}
where $\delta X^+(t)=\frac{1}{2}\left(\delta a^{\dagger}(t)+\delta a(t)\right)$ are amplitude fluctuations of the intracavity field. For simplicity we have chosen the phase of the intracavity field such that $\alpha =\alpha^*$. The steady state equation for the optical field is then: 
\begin{equation}
\alpha = \frac{\sqrt{2\kappa_{in}}\alpha_{in}}{\kappa - i\Delta }
\end{equation}
where the cavity detuning is also modified by the static displacement, $\Delta =\Delta^0+g\bar{x}$. The linearised equations of motion for the fluctuations are then:

\begin{eqnarray}
m_\text{eff}\left[\delta\ddot{ x}(t)+\Gamma_m\delta\dot{ x}(t)+\omega^2_m \delta x(t)\right] &=& 2\hbar g\alpha \left[ \delta X^+(t) +  \frac{\beta A}{\tau_{t}}  \left( H(t)e^{- \frac{t}{\tau_t}} \right)\ast \delta X^+(t) \right] \\ \nonumber
&&+ \sqrt{2\Gamma_m m_\text{eff} k_b T}\xi(t)\\
\delta\dot{ a}(t) &=& -\left(\kappa - i\Delta\right)\delta a(t) + ig\alpha\delta x(t) +  \sqrt{2\kappa_{in}}\delta a_{in}(t)
\end{eqnarray}

The photothermal integral in (\ref{eq:SF_x}) has been replaced by a convolution with a Heaviside function $H(t)$ to preserve causality. Transformation into the Fourier domain (i.e. $\mathcal{F}\{ \delta x(t)\}=\delta x$) yields:

\begin{eqnarray}
\delta x &=& 2\chi(\omega)\hbar g\alpha\delta X^+ \left[ 1 + \frac{\beta A}{1+i\omega\tau_t} \right] \label{eq:SF_dx} \\
\delta a &=& \frac{ig\alpha\delta x + \sqrt{2\kappa_{i}}\delta \alpha_{i} }{D(\omega)} \label{eq:SF_da}
\end{eqnarray}
which uses the Fourier identity $\mathcal{F}\{H(t)e^{t/\tau_t}\} = \frac{\tau_t}{1+i\tau_t\omega}$, reducing the functional form of the photothermal force to a low pass filter with a corner frequency defined by the characteristic thermalization time. For brevity the mechanical and optical transfer functions have been introduced, $\chi(\omega)^{-1}=m_\text{eff}\left(\omega^2_m-\omega^2+i\omega\Gamma_m\right)$ and $D(\omega) = \kappa+i(\omega-\Delta)$ respectively. From (\ref{eq:SF_dx}) it is clear both optomechanical effects arise from intracavity amplitude fluctuations $\delta X^+$ which can be expressed as:

\begin{align}
\delta X^{+} &= \frac{1}{2}\left( \delta a^\dagger(-\omega) + \delta a(\omega) \right) \\
 &= \frac{\alpha\delta x}{x_\text{ZPF} D(\omega)D^*(-\omega)} \bigg[-\Delta g_0 + \mathcal{O}\{\delta a_{in}\}\bigg]
 \end{align}

%
where the optomechanical coupling rate has been normalized by the oscillators' zero point motion ($x_\text{ZPF}$), that is $g_0=gx_\text{ZPF}$, and the fluctuations of the injected optical field have been grouped into a single term $\mathcal{O}\{\delta a_{in}\}$. In the case of a coherent optical drive this term quantifies the additional thermomechanical noise generated by optical vacuum fluctuations, commonly known as quantum backaction. Here, we discard the term $\mathcal{O}\{\delta a_{in}\}$ since dynamical instabilities and thermal noise greatly exceed quantum backaction. Substituting the resulting intracavity amplitude fluctuations into (\ref{eq:SF_dx}) gives:


\begin{eqnarray}
\delta x &=& \chi(\omega)\bigg\lbrace\frac{ -4 g_0^2 |\alpha|^2 \omega_m \delta x  m_\text{eff} \Delta}{D(\omega)D^*(-\omega)} \left[ 1 + \frac{\beta A}{1+i\omega\tau_t} \right] + \sqrt{2\Gamma_m m_\text{eff} k_b T}\bigg\rbrace\\ 
  &=& \chi'(\omega)\sqrt{2\Gamma_m m_\text{eff} k_b T} 
\end{eqnarray}

where we have made the substitutions $x^2_\text{ZPF} = \hbar/2m_\text{eff}\omega_m$. The modified mechanical susceptibility $\chi'(\omega)$ includes the radiation pressure and photothermal optomechanical coupling and can be written in the form $\chi'(\omega)^{-1} = m_\text{eff} (\omega_m^2 + 2 \omega \delta \omega_m - \omega^2 + i\omega[\Gamma_m + \delta\Gamma_m])$. Isolating real and imaginary terms gives the modification to the mechanical resonance frequency $\delta\omega_m$ and decay rate $\delta\Gamma_m$ as:

\begin{eqnarray}\label{eq:delomega} 
\delta\omega_m = \frac{A(\omega)}{2\omega_m}\left\lbrace 
\left[ (\kappa^2+\Delta^2-\omega^2)\left(1+\frac{\beta A}{1+\omega^2\tau_t^2}\right) - \frac{2\kappa\omega^2\tau_t\beta A}{1+\omega^2\tau_t^2}\right] 
\right\rbrace
\end{eqnarray}

\begin{eqnarray}
\label{eq:delgamma} 
\delta\Gamma_m  = - A(\omega) \left\lbrace \left[(\kappa^2+\Delta^2-\omega^2)\frac{\tau_t\beta A}{1+\omega^2\tau_t^2} + 2 \kappa \left(1+\frac{\beta A}{1+\omega^2\tau_t^2}\right)\right]
\right\rbrace
\end{eqnarray}

where $A(\omega) = \frac{4 g_0^2 |\alpha|^2 \omega_m \Delta}{|D(\omega)D^*(-\omega)|^2}$ is proportional to the magnitude of the cavity response.


Despite extensive experimental work on linear dispersive optomechanics over the last decade the photothermal effect has not received the same level of interest, potentially due to the perceived limitations of its inherently dissipative origin. To illustrate the nature of photothermal coupling we take the limit where it dominates over radiation pressure, namely $\beta A>>1$. In this situation two distinct regimes can be realised depending on the thermalization rate $\tau_t$. As previously mentioned the thermal response can be represented by a low pass filter with corner frequency $1/\tau_t$. If the corner frequency is larger than $\omega_m$ then the dominant effect is a shift in the mechanical resonance frequency. Conversely if the corner frequency is less than $\omega_m$ then the resulting delay in the photothermal force leads primarily to a modified mechanical decay rate. These two scenarios are analogous to the picture of ``unresolved'' and ``resolved'' dynamics in dispersive optomechanics respectively.

\section*{Experimental Details}

Our experiment is contained inside a sealed sample chamber which is mounted within an Oxford Instruments closed cycle helium-3 cryocooler with a base temperature of 300~mK. Inside the sample chamber the microtoroid is mounted on Attocube translation stages for linear positioning relative to the tapered optical fibre, which itself is mounted on a custom made glass taper-holder to match the thermal contraction/expansion during thermal cycling. Imaging of the microtoroid and taper is achieved via mounted microscopes directed through a set of windows fixed into the bottom of the cryostat. A stainless steel access tube connected to the sample chamber is used to introduce helium-4 gas into the sample chamber.

Vibrations from the continuously running pulse-tube cooler (PTC) are transmitted through the cryostat to the sample chamber, preventing stable positioning of the taper-toroid separation. To circumvent this issue, on-chip stabilization beams are added using photolithography during the fabrication of the microtoroid. With one beam located on either side of the microtoroid the taper may ``rest'' upon the beams, making the relative vibration common-mode and stabilizing the separation to a high precision. This makes it possible to indefinitely maintain critical coupling even while the PTC is operational. To minimize light scattering out of the taper mode the support beams are only $5~\mu$m wide and selectively thinned from $2~\mu$m to less than 500~nm. 

\begin{figure*}[t]
\begin{center}
\includegraphics[width=.9\textwidth]{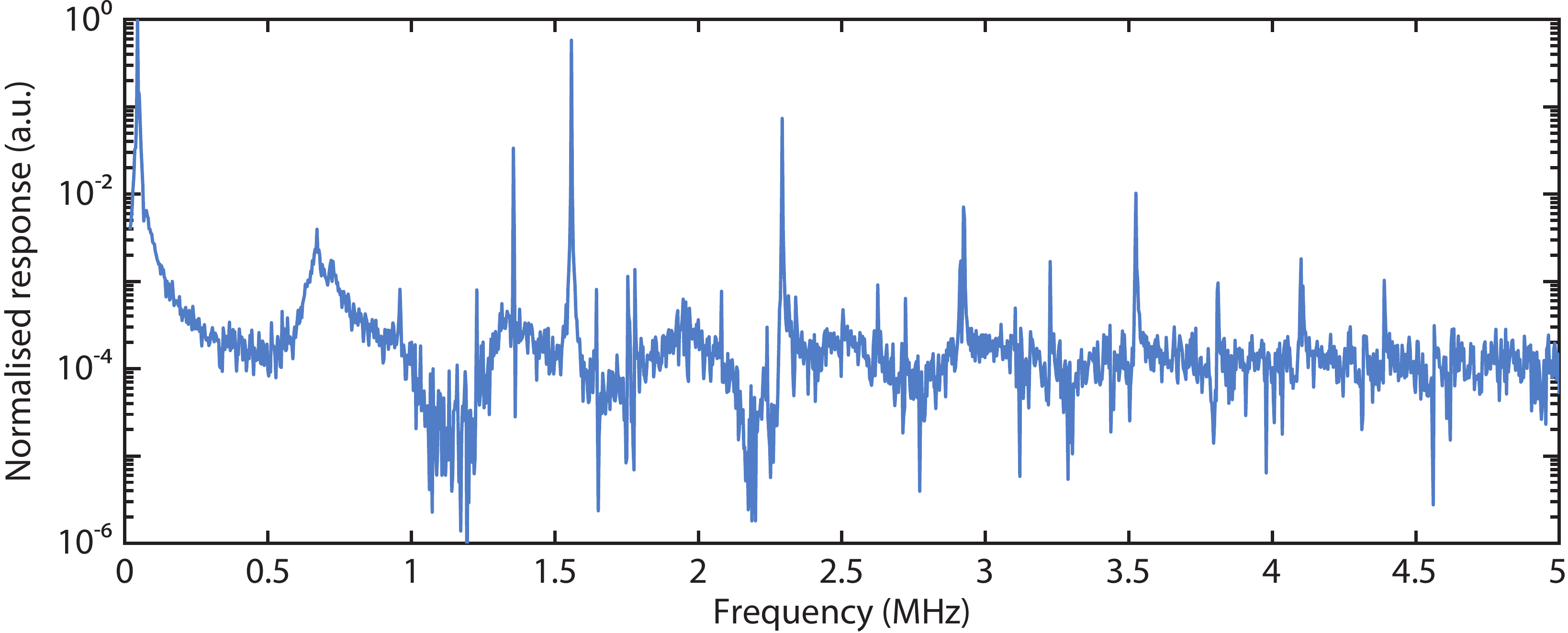}
\end{center}
\caption{\label{fig:NA} \textbf{Superfluid mode spectrum.}
Network analysis measurement showing the spectrum of third sound modes present in our resonator at 900~mK. Each peak in the spectrum corresponds to a different third sound mode. 
}
\end{figure*}

Ultrasensitive readout of the phase fluctuations imprinted by the motion of the superfluid film is achieved via homodyne detection using a fibre interferometer. The motion of the third sound waves was observed using both spectral and network analysis. Fig. \ref{fig:NA} shows a typical mode spectrum acquired via network analysis, showing the large range of third sound modes present in our system. To lock the relative phase angle between the local oscillator and signal we generate a 200~MHz amplitude modulation before the microtoroid. Mixing down the AC component of the  photocurrent at the modulation frequency then provides a phase dependent error signal for the interferometer that is filtered and applied to a piezoelectric fibre stretcher. In this configuration the DC component also provides a dispersive error signal for the cavity lock which is enhanced by the local oscillator, enabling stable operation even with nanoWatts of optical power in the signal arm of the interferometer.

\section*{Non-equilibrium bath coupling}

\begin{figure*}[t]
\begin{center}
\includegraphics[width=\textwidth]{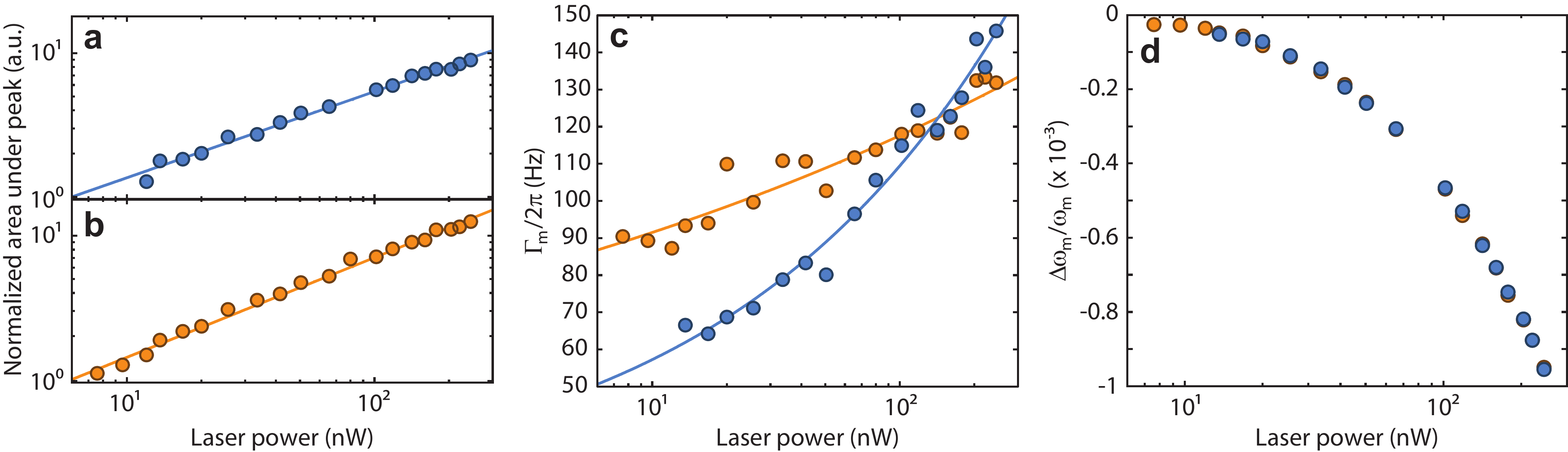}
\end{center}
\caption{\label{fig:onresonance} \textbf{On-resonance behaviour.}
Optical power dependence of the mechanical properties of a superfluid mode at $522.5$~kHz (orange circles) and another at $482$~kHz (blue circles), when the laser is locked to the cavity resonance. Measurements are performed with the laser coupled to a microtoroid optical mode at 1555.1~nm.
(\textbf{a, b}) Measured thermal energy of the superfluid mode versus input laser power. The blue line is a power-law fit to the 482~kHz mode with scaling $\propto P^{0.6}$. The orange line is the fit to the 522.5~kHz mode with scaling $\propto P^{0.69}$.
(\textbf{c}) Mechanical linewidth ($\Gamma_\text{m}/2\pi$) versus optical power. The blue line is a power-law fit given by $\Gamma_\text{m}/2\pi = 16.0\times P^{0.38}+19.3$ and the orange line is the fit $\Gamma_\text{m}/2\pi = 49.2\times P^{0.14}+23.4$. 
(\textbf{d}) Relative shift in mechanical resonance frequency versus laser power. This shift to lower frequencies as the laser power is increased is likely due to the density of the superfluid component decreasing as the film is heated, decreasing the third sound speed (see Eq. \ref{eq:thirdsound})\cite{Pramudya_thesis13}. 
}
\end{figure*}

As commented in the main Letter, it is well-known that dynamical backaction due to detuned laser driving in a cavity optomechanical system can allow heating or cooling of the mechanical motion. In this section, we investigate the effects of the optical probe on the superfluid modes with the laser locked to the cavity resonance, in the absence of dynamical backaction. Specifically, spectral analysis was performed on two superfluid modes with resonance frequencies of $522.5$~kHz and $482$~kHz, as the coupled laser power was varied from $7-250$~nW. This measurement allows modifications of the mechanical resonance frequency, dissipation rate, and mode energy to be measured as function of optical power. 

In these experiments we observe an increase in mode energy, equivalent to heating, of the superfluid mode. This increase in mode energy is characterized by a power law scaling of $P^{0.60}$ for the $482$~kHz mode and $P^{0.69}$ for the $522.5$~kHz mode (Figs. \ref{fig:onresonance}a, b). This non-integer power scaling of mode energy is markedly different to the linear power scaling expected if the temperature increase was due to quantum backaction or bulk heating of the superfluid. Concurrent to the increase in mode energy, it was observed that for both modes the mechanical linewidth increases with increasing laser power (Fig. \ref{fig:onresonance}c) and the mechanical resonance frequency decreases with increasing laser power (Fig. \ref{fig:onresonance}d). It should also be noted that we have observed behaviour that we attribute to boiling off the superfluid layer, that is, for high laser powers the superfluid modes rapidly broaden and increase in frequency until they are no longer observable. However this occurs at laser powers greater than $5$~$\mu$W, over an order of magnitude higher than the powers used in experiment. 

\begin{figure}[t]
\begin{center}
\includegraphics[width=0.6\textwidth]{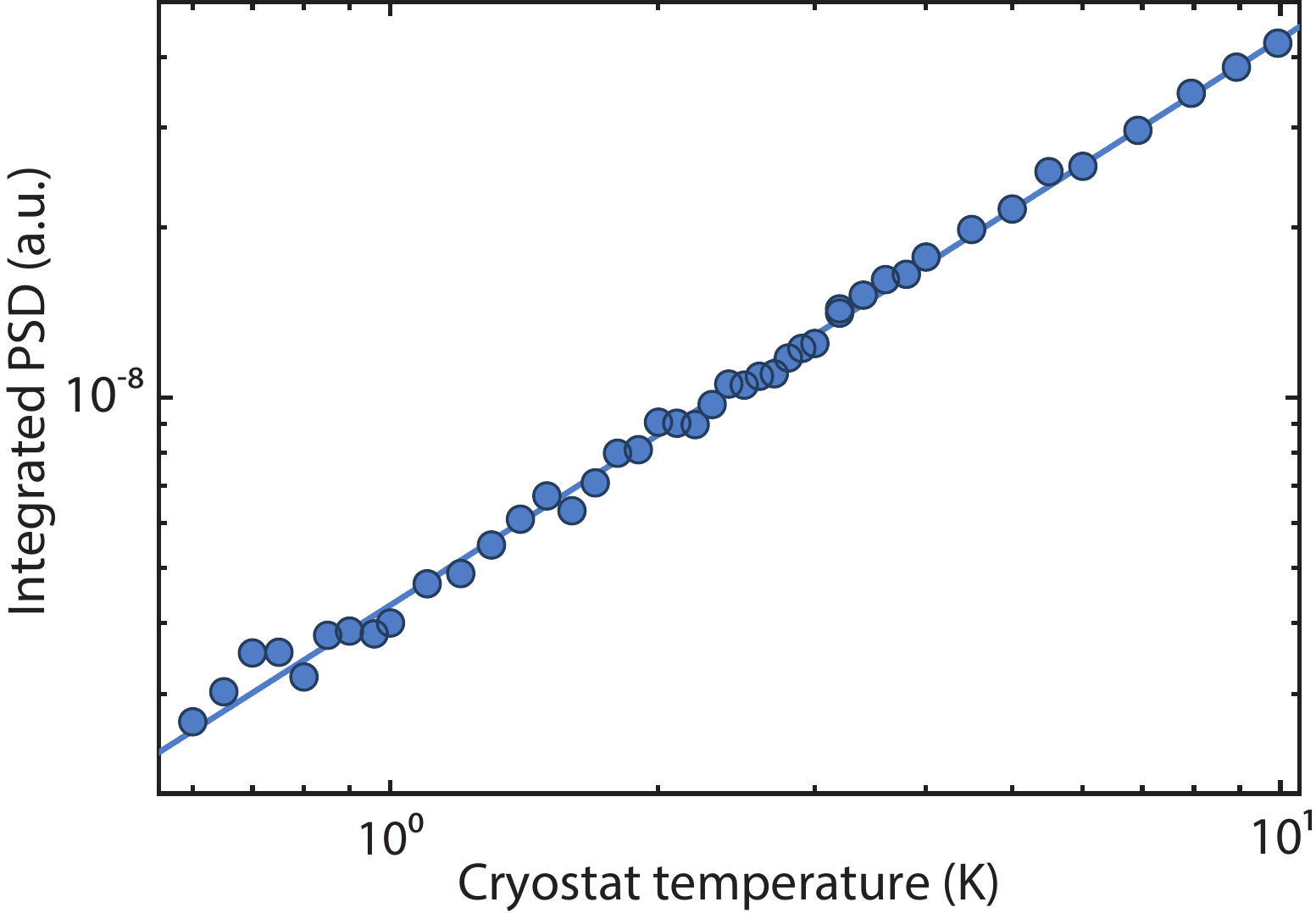}
\end{center}
\caption{\label{fig:thermometry} \textbf{Microtoroid thermomentry.}
The mode energy, proportional to the integrated power spectral density (PSD), of a microtoroid mechanical mode at $1.35$~MHz as the cryostat temperature is increased. The solid line is a linear fit to the experimental data showing there is good thermal anchoring of the microtoroid to the cryostat.}
\end{figure}

To further rule out bulk heating effects, measurements were performed on a mechanical mode of the microtoroid itself at $1.35$~MHz. In this case the thermal energy of the mechanical mode, dissipation rate, and resonance frequency were all found to be independent of the laser power when varied from 7~nW to 250~nW whilst on-resonance. After confirming that on-resonance probing does not heat the mechanical modes of the microtoroid, we varied the temperature of the cryostat from $10$~K to $600$~mK with fixed optical power (100~nW). From the linear fit to data in Fig. \ref{fig:thermometry} we found that the temperature of the microtoroid closely follows that of the cryostat.

To investigate the effect of coupling to a generalized non-equilibrium bath we first assign an effective temperature $T_\text{B}\left(\omega\right) = \frac{\hbar\omega}{k_B} \left[\text{ln}\left(\frac{S_{BB}(\omega)}{S_{BB}(-\omega)}\right)\right]^{-1}$ to the bath, where $S_{BB}$ is the corresponding noise spectral density\cite{Clerk_RMP10}. Since this non-equilibrium bath is independent to the thermal bath ($T_{th}\propto k_{B}T$), then the associated coupling rate is not given by the intrinsic decay rate ($\Gamma_{0}$), but by the generalized expression $\Gamma_\text{B} = \frac{x_\text{zpf}^2}{\hbar^2}\left(S_{BB}\left(\omega_m\right)-S_{BB}\left(-\omega_m\right)\right)$ \cite{Clerk_RMP10}. Importantly, we note that if the effective bath temperature is negative, namely $S_{BB}(-\omega)>S_{BB}(\omega)$, then the coupling rate is also negative. Taking the high quality factor limit gives the final temperature of the oscillator due to the presence of the thermal bath and the non-equilibrium bath:
\begin{equation}\label{eq:temp} 
T_\text{Final} = \frac{T\Gamma_0 + T_\text{B}\Gamma_{\text{B}}}{\Gamma_0 + \Gamma_{\text{B}}}.
\end{equation}
Depending on how the non-equilibrium bath is excited via the optical field, specifically the ratio of $S_{BB}(-\omega)$ to $S_{BB}(\omega)$, it is possible to obtain linewidth narrowing or broadening in concert with increased mode energy. Furthermore, it has been reported in Ref.~\cite{Meenehan_PRA14} that the bath temperature may depend nontrivially on the coupled laser power, resulting in a final mode temperature that permits the non-integer power law scaling behaviour we observe in Fig.~\ref{fig:onresonance}.